\documentclass[twocolumn]{aa}  

\usepackage{natbib}
\usepackage{amsmath}
\bibpunct{(}{)}{;}{a}{}{,} % to follow the A&A style

\usepackage{color}

\usepackage{graphicx}

\usepackage{txfonts}
\usepackage{hyperref}
\newcommand{\msun}{{\rm M}_\odot}

\newcommand{\teff}{T_{\rm eff}}
\newcommand{\tbb}{T_{\rm BB}}

\newcommand{\logg}{\log{g}}
\newcommand{\loghhe}{\log{\left(N_{\rm H}/N_{\rm He}\right)}}
\newcommand{\logche}{\log{\left(N_{\rm C}/N_{\rm He}\right)}}

\definecolor{azul}{rgb}{0,0,.8}
\definecolor{rojo}{rgb}{1,0,0}
\begin{document}

%\title{ \txa{Analysis of a group of DC white dwarfs proposed as spectrophotometric standards}}

%\title{ \txa{Analysis of a group of DC white dwarfs with nearly perfect black-body colors}}

\title{On the nature of black body stars}

\author{Aldo Serenelli\inst{1,2} \and 
 Ren\'e
  D. Rohrmann\inst{3} \and 
 Masataka Fukugita\inst{4,5}}
 
\institute{Institute of Space Sciences (ICE, CSIC) Campus
 UAB,
  Carrer de Can Magrans, s/n, E-08193, Bellaterra,
 Spain
  \\ \email{aldos@ice.csic.es} \and Institut d'Estudis Espacials de
  Catalunya (IEEC), 
 C/Gran Capita, 2-4, E-08034, Barcelona, Spain
  \and
 Instituto de Ciencias Astron\'omicas, de la Tierra y del
  Espacio (CONICET), 
 Av. Espa\~na 1512 (sur), 5400 San Juan,
  Argentina \and 
 Kavli Institute for the Physics and Mathematics of
  the Universe, University of Tokyo, Kashiwa 277-8583 Japan \and 
  Institute for Advanced Study, Princeton, 08540 NJ, USA}

\abstract{A selection of 17 stars in the Sloan Digital Sky Survey, previously identified as DC class white dwarfs (WDs), has been reported to show spectra very close to blackbody radiation in the wavelength range from ultraviolet to infrared. Due to the absence of lines and other details in their spectra, the surface gravity of these objects has not been previously well constrained and their effective temperatures have been determined by fits to the continuum spectrum using pure helium atmosphere models. We compute model atmospheres with pure helium and H/He mixtures and use Gaia DR2 parallaxes available for 16 out of the 17 selected stars to analyze their physical properties. We find that the atmospheres of the selected stars are very probably contaminated with a trace amount of hydrogen as $-6 \leq \loghhe \leq -5.4$. For the 16 stars with Gaia parallaxes, we calculate a mean stellar mass  $0.606\pm 0.076\,\msun$, which represents typical mass values and surface gravities ($7.8<\log g<8.3$) for WDs.}


\maketitle
%
%________________________________________________________________

\section{Introduction}\label{s:intro}

DC type white dwarfs (WDs) comprise degenerate stars showing only continuous spectra which are commonly atributed to helium-rich atmospheres with effective temperature ($\teff$) lower than $\approx 11000$~K.  Specifically, they have long been recognized as descendants of DB class stars (WDs with He I lines and no other elements present in their spectra) as they cool down below the spectral line detection \citep{baglin:1973}. It was also suggested by Baglin \& Vauclair that some DC WDs could be a result of the mixing in convective DA stars (almost pure H atmospheres). This process was proposed to explain the so-called non-DA gap in the range $5000$~K~$\la\teff\la6000$~K where few non-DA stars are found \citep{Bergeron:1997}. Using evolutive models and statistical analysis, \citet{Chen:2012} showed that such a deficit could originate from a combination of convective mixing and a higher cooling rate of the post-mixing WDs.
In addition, recent studies have showed a bifurcation of helium- and hydrogen-atmosphere cooling sequences on color-magnitude diagrams of Sloan Digital Sky Survey (SDSS) and Gaia passbands \citep{Gaia:2018,Kilic:2018} with both sequences coinciding below $\teff\approx7000$~K, where a high number of DC stars are confined. The observed split in the two sequences could be not only due to atmospheric composition, but also to an effect of the stellar mass distribution \citep{Kilic:2018}. Clearly, further work is necessary to understand the role of DC stars in the chemical evolution of cool WD atmospheres.

Recently, \citet[hereafter SF18]{sf18} have identified a group of stars that exhibit almost perfect blackbody spectra with no apparent absorption lines, which SF18 referred to as blackbody stars and to which we refer here simply as the SF18 sample or SF18 stars. Their study included mainly spectrophotometric data of the SDSS in DR7, but was also supplemented with ultraviolet photometry of Galaxy Evolution Explorer (GALEX) and infrared data of the Wide-field Infrared Survey Explorer (WISE). 
The selected group of objects, composed by 17 stars that mimic the blackbody emission, were previously classified as DC WDs in a number of studies based on spectral, photometric and kinematics analyses \citep{Kleinman:2004,Eisenstein:2006,Kleinman:2013,Kepler:2015,Gentile:2015}.
% (see also the Montreal White Dwarf Database\footnote{\href{http://dev.montrealwhitedwarfdatabase.org/}{http://dev.montrealwhitedwarfdatabase.org/}}). 
The potential interest of this sample  
relies on the simplicity of their spectral energy distributions, which makes them excellent objects for calibration purposes of photometric (SF18) and spectroscopic surveys \citep{lan:2018, narayan:2018}.
Deviations of the SF18 stars from blackbody colors from infrared to UV are really minuscule. The lack of spectral features prevents an accurate evaluation of the surface gravity ($g$), and previous effective temperature evaluations have been based on photometric and continuum spectrum fits with pure helium models. However, it is particularly unclear whether the continuum spectrum of pure-He atmosphere white dwarfs can lead to a spectral energy distribution that mimic so well that of a black body spectrum in the whole temperature range covered by the SF18 sample. 
A study of these stars, also, could provide further insight about the chemical evolution of cool WDs. Furthermore, this subpopulation of DC stars could also open a novel possibility of testing our understanding of cool, high-density stellar atmospheres.

In this work we study the stellar parameters of the SF18 sample by considering the spectral properties of white dwarf model atmospheres computed with different assumptions regarding their composition. We find that pure-He atmosphere models produce a blackbody spectrum that matches the observed sample of DC stars, but require  high surface gravities for the cooler stars. Atmospheres dominated by helium that contain a trace amount of hydrogen can also reproduce the data very well,  with $\log~g$ values more typical for WDs. The effective temperature of the models that fit best these stars differs, however, from the estimated blackbody temperature ($\tbb$) due to opacity effects. The difference between these temperatures depends on the blackbody temperature as well as on the amount of hydrogen pollution of the atmosphere. The inclusion of the Gaia data allows us to determine the physical parameters of these stars, their temperature, radius, mass and surface gravity, with the aid of atmosphere models, to a good precision, removing the ambiguity present when only photometry is avaliable.  We find the mass distribution of this sub-group of DC stars to be similar to that of (hotter) DB stars. This lends us to provide support that this sample of DC stars are cooler descendants of DBs.

The layout of the paper is as follows. In Section~\ref{s:models} we briefly review the model atmospheres used in this work. In Section~\ref{s:results} we show that models with pure-He composition or very low hydrogen pollution reproduce the blackbody properties of the SF18 stars, and in Section~\ref{s:parameters} we determine the stellar parameters of the stars. In Section~\ref{s:discussion} we discuss our results in the context of the evolution of DB and DC stars, as well as the potential use of the SF18 sample as calibrators for both photometric and spectroscopic surveys. Finally a summary and conclusions are presented in Section~\ref{s:conclusions}.

\section{WD atmosphere models} \label{s:models}

The white dwarf model atmospheres used in this work have been
computed within the assumption of plane-parallel geometry, LTE,
radiative-convective and hydrostatic equilibrium. Convective energy
transport is treated in the mixing-length
approximation (ML2, e.g. \citealt{salaris:2008}). Particle populations for a mixture of hydrogen and
helium (H, H$_2$, H$^+$, H$^-$, H$_2^+$, H$_3^+$, He, He$^-$, He$^+$,
He$^{++}$, He$_2^+$, HeH$^+$, and e$^-$) are derived from the
occupation probability formalism. All relevant radiative opacities are
considered. Details of the code can be found in \citet{rohrmann:2012}
and references therein. We use two classes of models according to
composition: pure-He, and He-dominated atmospheres with  H abundances
$\loghhe=-2,...,-8.5$. Results for pure hydrogen atmospheres are also
considered but only for reference.

\section{SF18 stars in the color-color plane} \label{s:results}

\begin{figure}
\includegraphics[width=.48\textwidth]{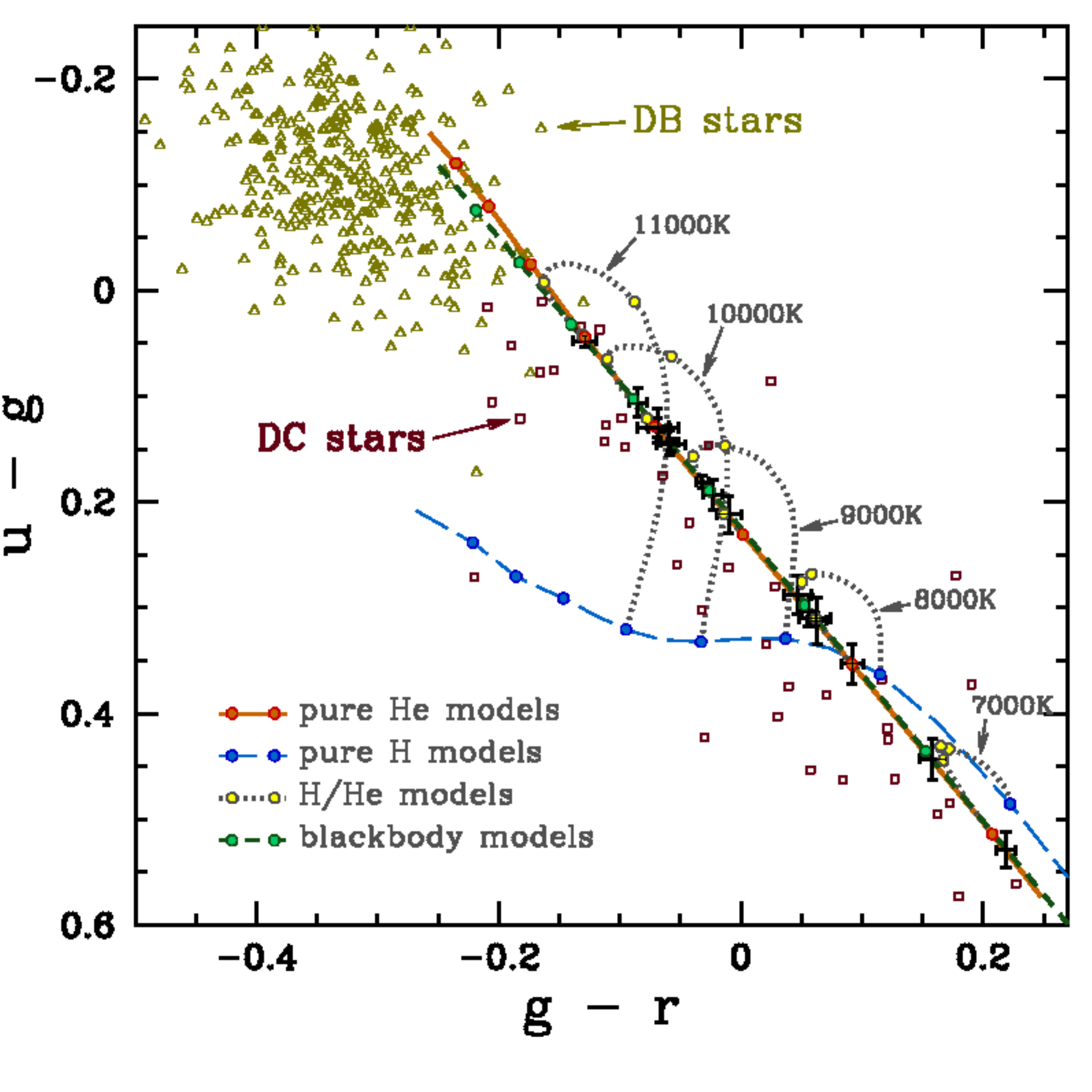}
\caption{Color-color diagram for a sample of white dwarf stars from the SDSS DR7 catalog \citep{Kleinman:2013}. The SF18 sample is shown with error bars. Lines denote theoretical color-color curves for blackbody (green short-dashed), pure-He (red solid), pure-H (blue long-dashed), and mixed H/He (dotted) atmosphere models as indicated in the figure legend, in the range 14000-7000~K. Small circles on top of the curves showing blackbody, He atmosphere and H atmosphere models with marks indicating 1000~K intervals. Those on the bridging between H and He  models mark the $\loghhe=-2, -4, -6$ values.  \label{fig:color-color}}
\end{figure}

We present in Figure~\ref{fig:color-color} the $(u-g)$ vs
$(g-r)$ color-color diagram for white dwarfs from the SDSS DR7 white
dwarf catalog \citep{Kleinman:2013}, where $u,g,r$ are the SDSS color
bands \citep{fukugita:1996}. We include DB (\ion{He}{1} 4471 \AA) and
DC (featureless spectrum) with the spectroscopic classification taken
from the original catalog. Randomly selected samples are shown for
each class to avoid overcrowding the plot.  The 17 stars of the SF18 sample
are shown as black data points with error bars.

Figure~\ref{fig:color-color} includes theoretical predictions for
different classes of white dwarf atmosphere models characterized by
their composition. The orange solid line represents the predictions of
pure-He atmosphere models for varying $\teff$ from about 14000~K
(upper left mark) down to 7000~K (lower right mark) with a 1000~K
step. Pure-H atmospheres have colors very different from a blackbody
spectrum in the temperature range of interest. These models are shown
with blue long-dashed lines in Fig.\,\ref{fig:color-color}. Colors of
atmospheres composed of H/He mixtures vary with the relative abundance
of H with respect to He. The dotted lines in the figure denote
sequences of models with constant $\teff$ and decreasing amount of
hydrogen going from pure-H to pure-He model atmospheres. Here, yellow
circle ticks denote $\loghhe= -2, -4$, and -6. When the H/He ratio
decreases to $\loghhe \lesssim -6$, they approach close to the pure-He
models, within the photometric uncertainties of the SF18 sample. 
It is to be noted that for $\loghhe \lesssim -6$
hydrogen lines remain hidden in helium-rich atmospheres; the
equivalent width of H$\alpha$ drops below $\sim 0.2\,\hbox{\AA}$
\citep{weidemann:1995}. All models in the figure correspond to
$\logg=8$: 
 $(u-g)$ and $(g-r)$ colors from pure-He atmospheres
depend very weakly on $\logg$. Finally, the green short-dashed line
represents results for blackbody spectra. Circle ticks represent
temperature intervals of 1000~K.

Note that the SF18 sample falls
exactly on top of the blackbody predictions. This is the result of
the selection of stars done by SF18, in which a star with a spectrum
that deviated substantially from a blackbody, was discarded. All
stars in the sample have a blackbody temperature $\tbb$ between 7000~K and
12,000~K, defined as the temperature of a blackbody that reproduces
all observed colors. In this temperature range, colors from pure-He
models almost perfectly overlap with blackbody colors. This is also
true for models with $\loghhe = -6$. At hotter ($g-r<-0.15$)  temperatures 
pure-He model predictions start to deviate from the blackbody
results. The same is true for the cooler temperatures ($g-r > 0.25$), as shown below in Fig.~\ref{fig:galex}.
These results support the identification of
stars in the sample as  DC stars either with pure-He envelopes or
with very small amount of H. On the other hand, it is evident
from the figure that DC stars do not always display colors that
match those of a blackbody and, in this regard, the SF18 sample 
represents a peculiar minority of the whole DC population. Note that
DC is a label of observational classification that implies the absence of spectral
features. While this is usually understood as due to He-dominated
atmospheres, it includes stars with trace abundances of other chemical elements with 
abundances low enough to not show spectral lines but large enough to affect the shape of the continuum.

It is possible that the SF18 sample is formed by DQ stars with very small traces of carbon, too
low to have any discernible feature in SDSS
spectra. \citet{Koester:2006} have shown that in C/He atmospheres the
effect of carbon in the $(u-g)$ vs $(g-r)$ plane vanishes for $\logche
< -8$ in the temperature range of the SF18 stars. Spectroscopically, the
lowest carbon abundance measured in DQ stars is $\logche \approx -7.5$
\citep{Koester:2006,Kepler:2015}. On the other hand, evolutionary
models \citep{camisassa:2017} indicate that stars with initially
pure-He envelopes undergo a rapid enrichment of carbon due to dredge
up as stars cool down from $\teff \approx 12500~K$ to 7000~K. The
carbon abundance then increases up to $\logche \approx -3$ for white
dwarf models of different stellar mass. However, in order for the SF18 stars to have a C 
abundance below the detectability level but large enough to affect colors, a finely tuned competition between dredge up and gravitational sedimentation has to occur throughout this effective temperature range. This possibility is made even less probable by the fact that the sample includes stars of similar mass and effective temperatures that differ by a few thousand degrees, a range over which models show much larger variations in surface C abundances than allowed by detectability limits.

The comparison between
models and data is more revealing when UV bands are included.
Figure~\ref{fig:galex} compares the SF18 sample with our models in
the Galex  
(FUV-NUV) vs $(g-r)$ plane. All stars are
perfectly matched to the blackbody color due to their selection criterion in SF18, which 
included UV photometry. 
Pure-He models also provide a good description of the data for stars bluer than $(g-r)=0.1$. For
cooler stars with $(g-r) > 0.1$, however, the $\logg=8$ canonical pure-He model
deviates from the blackbody results, and does not
reproduce, in particular, the three coolest stars. UV colors from
pure-He models show a larger sensitivity to $\logg$ than optical
colors, and models with $\logg=9$ remain close to blackbody colors at
cooler temperatures. This would suggest that the cool stars could
be quite massive white dwarfs with masses around 1.2~$\msun$, i.e. in
the realm of WDs with ONe or CO-Ne hydrid cores rather than typical CO
cores \citep{althaus:2005,doherty:2017}. A more satisfactory result is
found with the H/He models with $\loghhe=-5.4, -6$ also shown in
Figure~\ref{fig:galex}. These models at $\logg=8$ reproduce colors 
of the SF18 sample better than pure-He models across the whole range,
corresponding to typical WD masses of about 0.6~$\msun$ in agreement
with the mean mass value for DBs (e.g. \citealp{koester:2015}). We come back to this in the next section, in which we determine stellar parameters using the recent
astrometric results of Gaia DR2 \citep{GaiaDR2}.

\begin{figure}
\includegraphics[width=.48\textwidth]{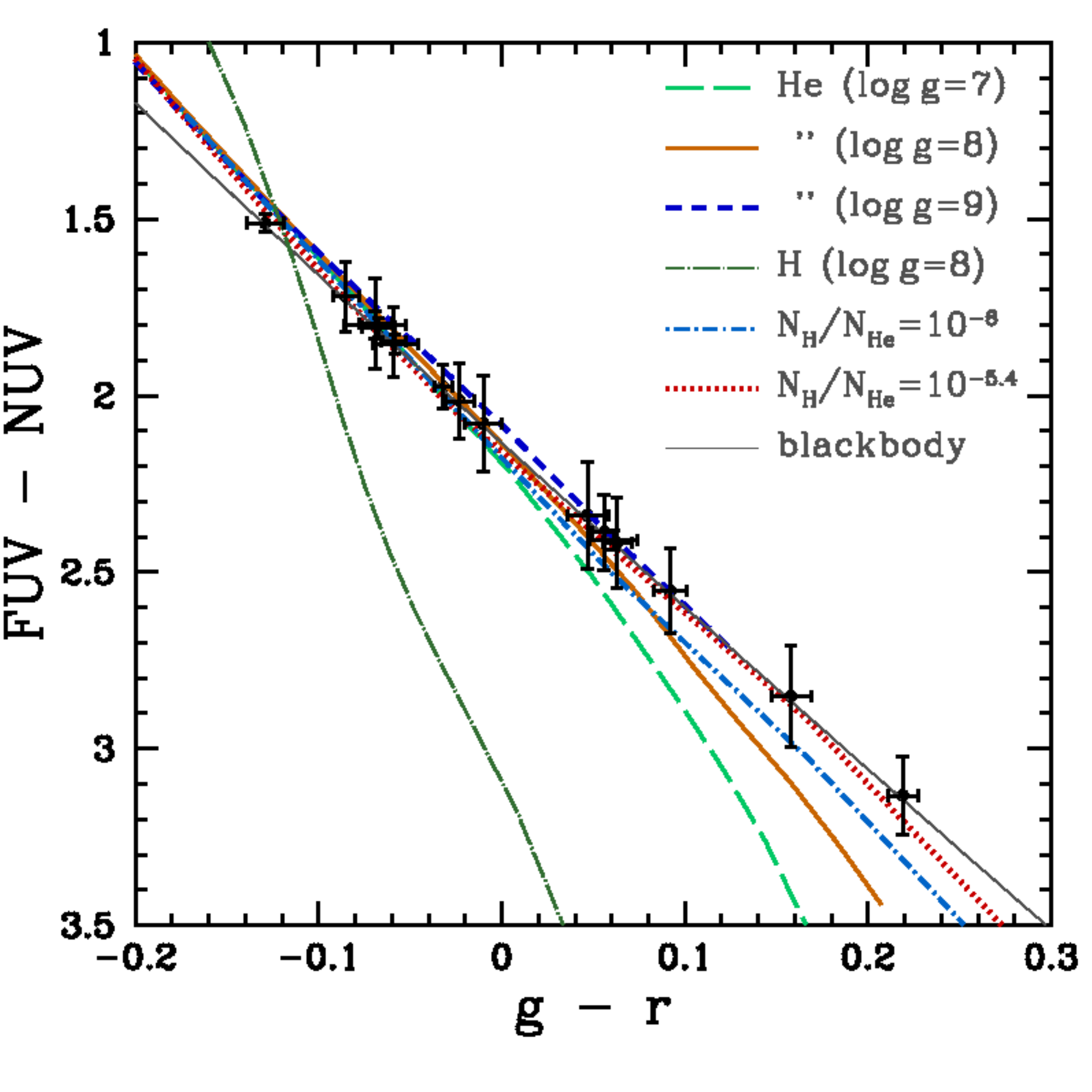}
\caption{(FUV-NUV)-$(g-r)$ color-color diagram. Stars in the SF18 sample are shown as black dots with error bars. Lines denote theoretical models as indicated in the legend. Models with H/He mixtures correspond to $\logg=8$. The H/He models with $\loghhe=-5.4$ reproduce all data of stars in the SF18 sample. \label{fig:galex}}
\end{figure}

\section{Determination of physical parameters} \label{s:parameters}

The physical parameters of the stars, in particular radius and mass, can be determined with the aid of model atmospheres, provided their distances are known. The recent Gaia DR2 \citep{GaiaDR2} includes astrometric solutions for 16 out of the 17 SF18 stars. We have queried the Gaia DR2 catalog using CosmoHub
\citep{carretero:2017}. Parallaxes range between 4.5 and 14~mas,
i.e. distances between 70 and 230~pc, and errors between 1\% and 10\% with a mean of 4.5\%, not far
from end-of-mission expectations. Parallaxes and distances are listed
in Table~\ref{tab:results}. 

In order to determine the stellar radius, we consider the relation, valid for SDSS magnitudes:
\begin{equation}
m_b = -2.5 \log{\left[\pi R^2 \varpi^2\right]} - 2.5 \log{\left[\frac{\int{F_\nu S_{\nu,b} d\nu}}{\int{S_{\nu,b} d\nu}}\right]} - 48.60, \label{eq:funda}
\end{equation}
where $m_b$ is the observed magnitude in a given band $b$, $\varpi$ is
the parallax, R the radius, $F_\nu$ the astrophysical flux, and
$S_{\nu,b}$ the passband transmission in band $b$. From the observed colors, $\tbb$ can be derived straightforwardly. But in the relation above, $F_\nu$ depends on the unknown $\teff$. Therefore, model atmospheres are required to relate the observationally determined $\tbb$ with $\teff$ which, in turn, is used to determine $F_\nu$.  

\begin{figure}
\includegraphics[width=0.48\textwidth]{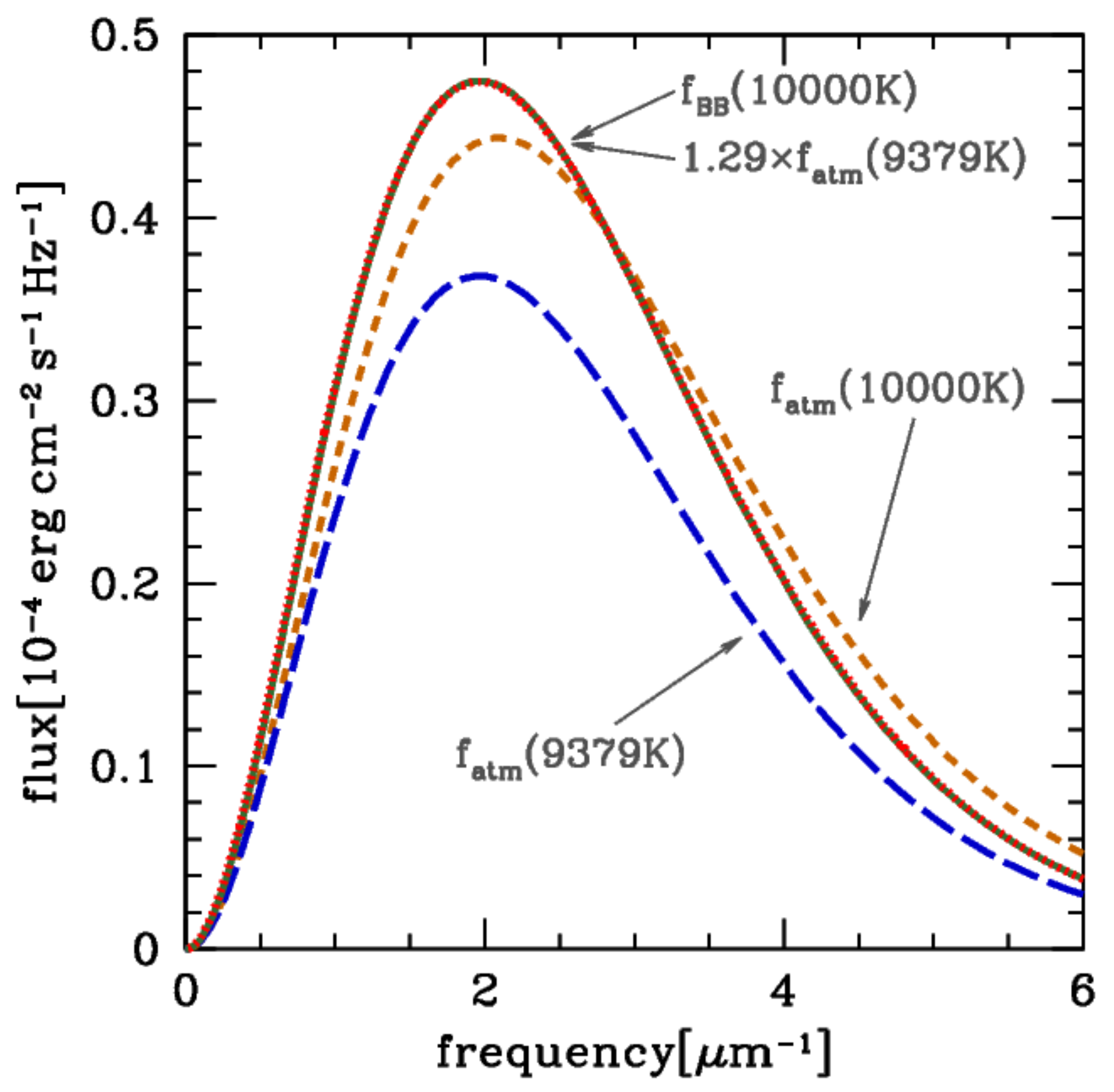}
\caption{Emerging monochromatic  flux. Orange short dashed line: pure-He atmosphere with $\teff = 10000\,\hbox{K}$; solid green line: blackbody spectrum with $\teff = 10000\,\hbox{K}$; blue long dashed line: pure-He atmosphere with  $\teff=9379\,\hbox{K}$; red dotted line: pure-He atmosphere with  $\teff=9379\,\hbox{K}$ scaled up by a factor 1.29 to match the total flux of a $\teff = 10000\,\hbox{K}$ spectrum. Model atmospheres correspond to $\logg=8$. \label{fig:bbspec}}
\end{figure}

The relation between $\tbb$ and $\teff$ depends on the physical conditions of the stellar atmosphere. 
To illustrate this, we compare in Figure~\ref{fig:bbspec}  the emerging flux of a
$\teff=10000\,\hbox{K}$ ($\logg=8$) pure-He atmosphere with a
blackbody spectrum also at $\teff=10000\,\hbox{K}$, i.e. both models have the same flux
$f(\teff)= \sigma \teff^4$, where $\sigma$ is the
Stefan-Boltzmann constant. The spectrum of the pure-He model is
slightly shifted towards the bluer side, compared to the black body
spectrum of the same $\teff$. The figure also shows the spectrum of a
pure-He model at $\teff=9379\,\hbox{K}$, which mimics the shape of a 
$\teff=10000$~K blackbody spectrum and hence its colors, but differ in the
flux normalization, which is smaller by a factor
$\left(9379/10000\right)^4=0.774$. 
%Similarly, the shape of the emerging spectrum of pure-He atmospheres is very close to a Planck function from about $\teff=12000~K$ down to 8000~K or lower (for higher $\logg$ values) but corresponding to a different higher blackbody temperature $\tbb$ with $\tbb - \teff= 500$-$800\,$K for $\logg=8$. This is discussed in more detail later in this section.

\begin{figure}
\includegraphics[width=0.48\textwidth]{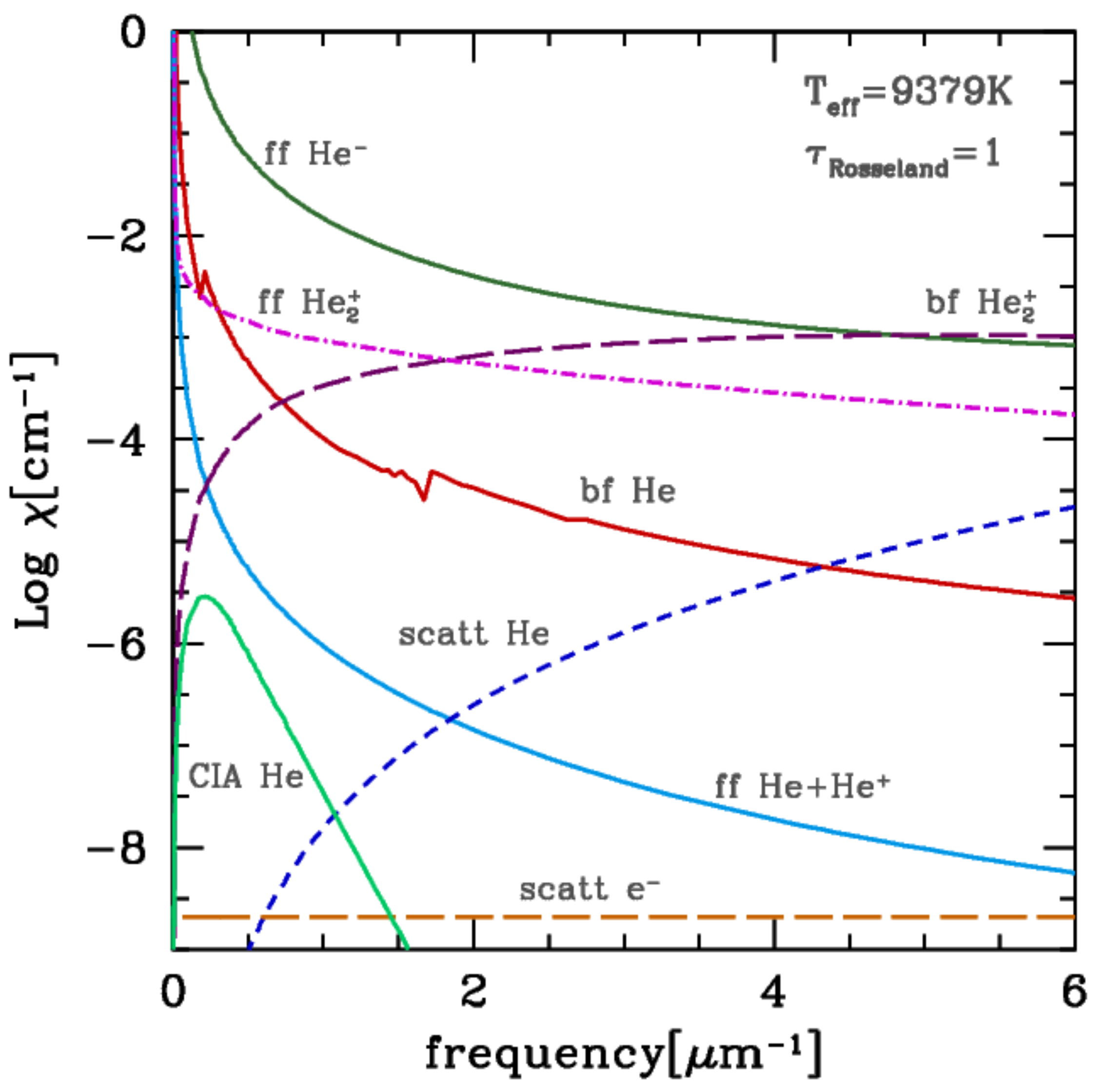}
\caption{Individual opacity contributions at $\tau_{\rm Ross}=1$ in a cool pure-He white dwarf atmosphere. Dominant contributions are ff processes of He$^-$ and He$_2^+$ and bf processes of He$_2^+$. Note that ff contribution from He$^-$ increases markedly at low frequencies and is responsible for shifting the emerging spectrum towards higher frequencies making it appear hotter, i.e. $\tbb > \teff$. \label{fig:opac}}
\end{figure}

The blueward shift of the spectrum of pure-He models with respect to a blackbody of
the same $\teff$ can be understood by looking at
Figure~\ref{fig:opac}, which shows the different contributions to the
opacity of the model with $\teff=9379\,\hbox{K}$ at optical depth
$\tau_{\rm Ross}=1$. Opacity is dominated by  He$^-$ free-free (ff)
processes and by bound-free (bf) processes of He$_2^+$ as the
secondary agent. These opacity sources have a relatively weak
dependence with frequency across most of the spectrum up to the dominant
ff He$^-$ contribution that increases towards longer wavelengths (see
Fig.\,\ref{fig:bbspec}). Increased flux blocking at longer wavelengths
shifts the spectrum towards the blue, giving it the shape of a
blackbody of higher temperature at the same integrated flux. In the
model shown in Fig.\,\ref{fig:bbspec}, bf transitions in He$_2^+$
become the single most important opacity source for frequencies larger
than $5\,\mu \hbox{m}^{-1}$, and increase for higher
frequencies. Therefore, it is expected that colors involving UV bands
will show deviations from blackbody atmospheres when bf He$_2^+$
processes become dominant. An increase in the number of free electrons that form  He$^-$ 
increases, e.g. at higher densities -larger surface gravities- or by the presence of trace hydrogen, will cause an increase of the ff He$^-$ opacity, keeping the total opacity closer to grey opacity. This is relevant to understanding the impact of gravity or hydrogen pollution in He-dominated atmospheres and it is the main reason behind results in Figure~\ref{fig:galex}.

\begin{figure}
\includegraphics[width=0.48\textwidth]{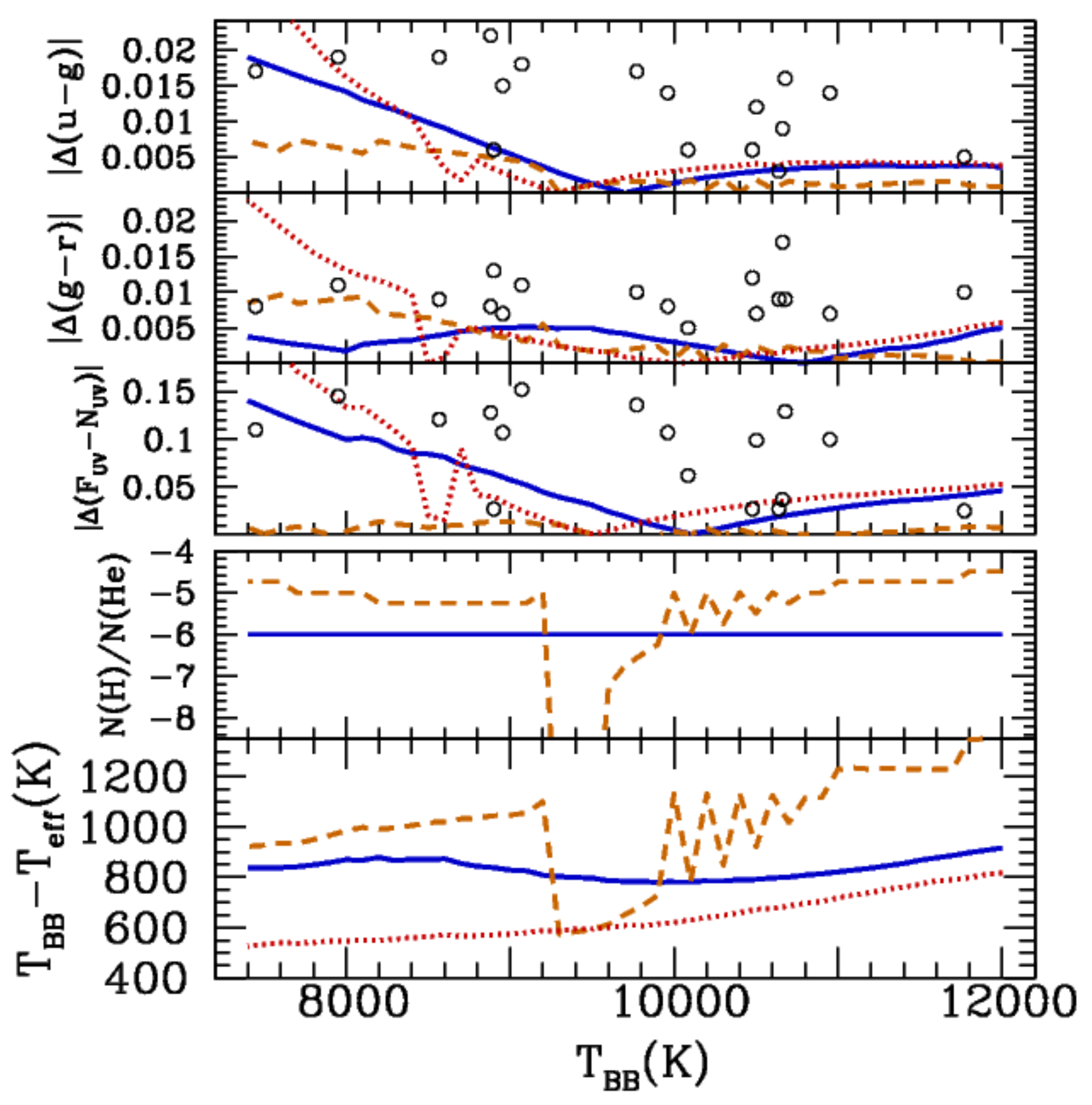}
\caption{Top three panels: absolute difference of colors between
  blackbody spectrum and pure-He atmospheres (red-dotted),
  $\loghhe=-6$ models (solid blue), and best-fit model (dashed
  orange). Circles denote residuals for the SF18 stars. 
 Fourth panel
  from top: $\loghhe$ values for the best fit model (dashed orange)
  and $\loghhe=-6$ (solid blue, constant). 
 Lower panel:
  $\tbb-\teff$ as a function of blackbody temperature and hydrogen
  abundance for pure-He, $\loghhe=-6$, and best-fit
  models. \label{fig:deltatemp}}
\end{figure}

Figure\,\ref{fig:deltatemp} shows in the top three panels the color difference between a blackbody and the 
model atmospheres that best adjust colors for different $\loghhe$ values. 
Results for pure-He models are shown in red dotted lines, for models with $\loghhe=-6$ in solid blue lines and
for the best-fit models with varying $\loghhe$ in dashed
orange. Differences are shown as a function of the inferred $\tbb$
(color temperature) value.  Circles denote residuals of the best
blackbody fit to the SF18 stars. The $\loghhe$ values of the best
fit models are shown in the fourth panel, where the curve for $\loghhe=-6$
is given for reference. Here we note that the best-fit models have in
most cases $\loghhe=-5.4$, somewhat above our adopted formal
spectroscopic detection limit. However, it is apparent that the models
with lower $\loghhe$ values also reproduce the blackbody colors with
residuals smaller than the differences between colors of the SF18 stars
and a black body spectrum. 

We note that the
temperature for the SF18 stars may be somewhat lower, by 
$100E(B-V))/0.01$~K, when extinction to these local white dwarfs is
taken into account (SF18), while $E(B-V)$ is typically 0.02 mag/(100pc)
in the solar neighbourhood.
It is difficult to determine extinction for individual WDs 
from photometry, since extinction is nearly parallel to the change of
temperature. Extinction hardly affects the extent of the deviation
of stars from these stars from the blackbody curve.

\begin{figure}
\includegraphics[width=.48\textwidth]{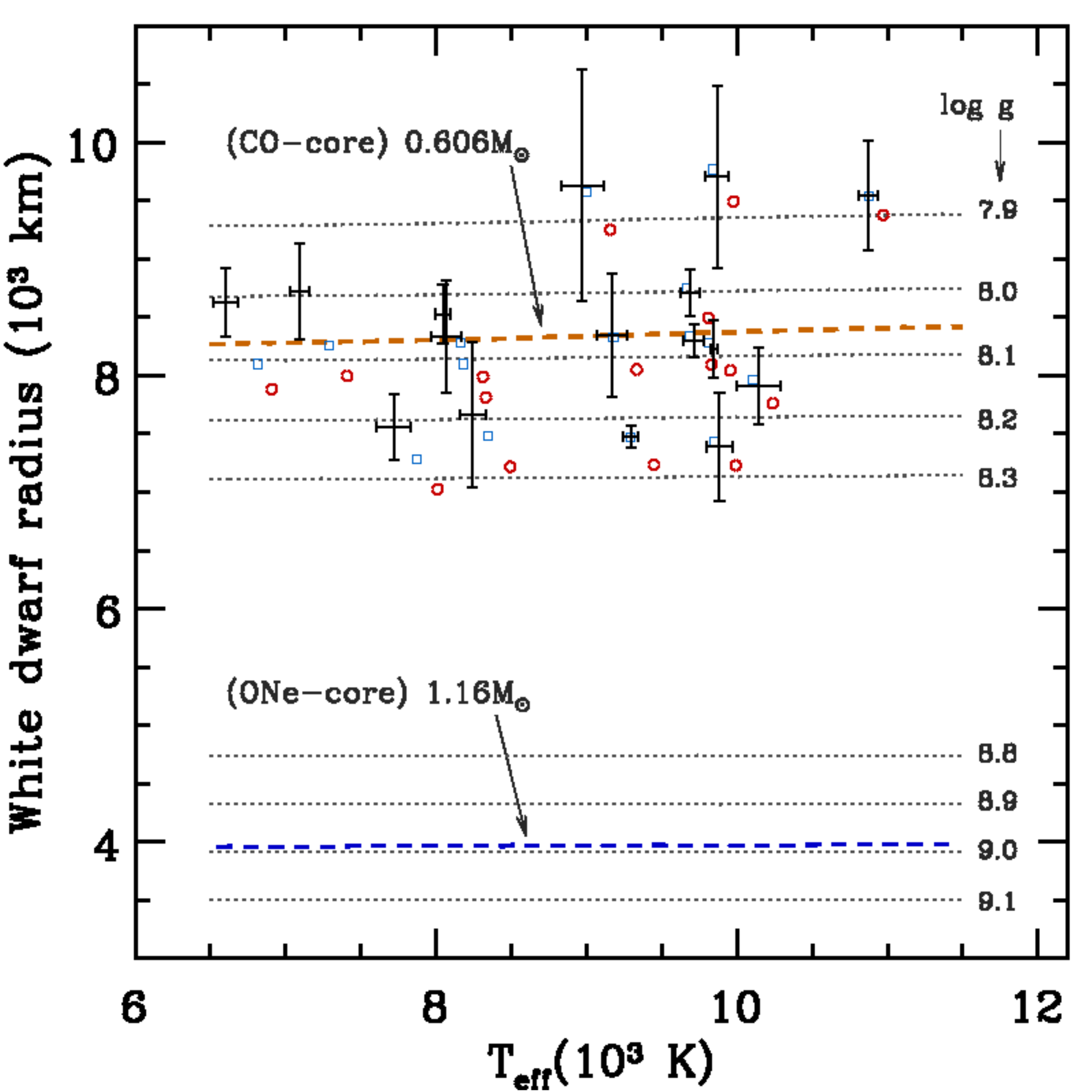}
\caption{$\teff$ and radius of stars in the SF18 sample with measured Gaia parallaxes. Results for different sets of model atmospheres are shown as: red circles ($\logg=8$, He-pure models), blue squares ($\logg=9$, pure-He models), black dots with error bars ($\logg=8$, $\loghhe=-6$). Overlaid with long-dashed lines are theoretical WD evolutionary tracks for 0.606~$\msun$ (CO-core) and 1.16~$\msun$ (ONe-core) WD models.  Short dashed lines show curves of constant $\logg$ determined from evolutionary models. \label{fig:radius}}
\end{figure}

In order to determine $R$ from Eq.\,\ref{eq:funda} we have used 
three different sets of model atmospheres:  1) He-pure with $\logg=8$; 2) He-pure with
$\logg=9$; 3) $\loghhe=-6$ and $\logg=8$. For each case 
the difference $\tbb-\teff$ can be approximated
analytically to better than 30~K as: 
\begin{equation}
\tbb-\teff = \begin{cases}
625 + 0.07 x + 1.41\times 10^{-5} x^2 \\
780 + 0.07 x \\
817 - 0.023 x + 1.15\time 10^{-5} x^2,
\end{cases}
\end{equation}
where $x = \tbb - 10000$. Results are shown in Figure~\ref{fig:radius}, where red circles, blue squares and black
crosses correspond to $\teff$ and radius estimates based on cases 1,
2, and 3,  respectively. For a given star, the
different radii reflect the different $\teff$ estimates through the
relation $R \teff^2 = C$, where $C$ is a constant. This simple relation
between $\teff$ and $R$ is the result of the shape of $F_\nu$ being
the same for all models that reproduce the colors of the star (see
Section~\ref{s:models}), so that the second term in Eq.\,\ref{eq:funda}
only depends on the normalization of $F_\nu$, given by $\teff^4$. The
figure also shows the evolutionary track of a CO-core 0.606~$\msun$
DB model, corresponding to $\langle M_{\rm DB}\rangle$
\citep{koester:2015}, and a track of an ONe-core 1.16~$\msun$ WD model. In
addition, curves of constant gravity determined from evolutionary
tracks are also shown for the two regimes. CO-core models are from
\citet{camisassa:2017} and ONe-core models from
\citet{althaus:2005}. 
 Error bars, shown for one case only, are determined using photometric uncertainties from SF18, Gaia DR2
parallax errors and $\teff$ errors from the uncertainties in fitting
$(u-g)$, $(g-r)$, $(FUV-NUV)$ colors.

Results in Fig.\,\ref{fig:radius} show that, regardless of the assumptions underlying the model atmospheres used, the radii of all the stars are consistent with them being CO white dwarfs, ruling out that they are massive WDs. 
Radii estimated with $\logg=8$ models and $\loghhe=-6$ are systematically larger than those obtained using He-pure models with the same $\logg$, and the difference increases towards lower temperatures. This reflects that, for given $\tbb$,  $\teff$  is lower for $\loghhe=-6$ models (bottom panel in Fig.\,\ref{fig:deltatemp}). 

Our results show that, by combining photometric and astrometric data, observational properties of stars in the SF18 sample are well described by $\logg=8$ and $\loghhe=-6$ WD atmosphere models. We therefore use stellar radii determined using these models as our reference values for the determination of further properties, $\logg$ and mass, of the SF18 stars. Our derivations make use of the DB evolutionary tracks from \citet{camisassa:2017}. Final results are reported in Table~\ref{tab:results}.

\begin{table*}
\caption{Physical parameters of the 16 stars in the SF18 sample with Gaia DR2 parallax measurements. Results correspond to models with $\loghhe=-6$, $\logg=8$ \label{tab:results}}
\setlength{\tabcolsep}{3pt}
%\begin{small}
\begin{tabular}{lcccccccccccc}  \hline 
\multicolumn{1}{c}{Star} &  $\varpi$ [mas] & Distance$^{\dagger}$ [pc] &$\teff$ [K]&  R [km] &  $\logg$ & M [M$_\odot$]  \\ \hline
J002739.497-001741.93 &   $4.366 \pm 0.344$ & $229  \pm 19 $ &    $9869\pm75$ &  $9705 \pm 780$& $7.84\pm0.09$ & $0.491\pm0.082$ \\
J004830.324+001752.80 &   $7.315\pm0.218$  & $137  \pm 4 $ &    $9847\pm25$  & $8224\pm250$ & $8.09\pm0.03$ & $0.623\pm0.026$  \\
J014618.898-005150.51 & $5.542\pm0.266$ &$180  \pm 9$ &    $10873\pm60$  & $9547\pm471$ & $7.87\pm0.06$  & $0.508\pm0.050$ \\
J022936.715-004113.63 &  $6.438\pm0.334$  & $155  \pm 8$  & $8066\pm100$ & $8332\pm480$ & $8.06\pm0.06$ & $0.603\pm0.051$ \\ 
J083226.568+370955.48 & $8.457\pm0.361$ & $118  \pm 5 $ &  $7094\pm65$ & $8721\pm411$ & $7.99\pm0.05$ & $ 0.561\pm0.043$ \\
J083736.557+542758.64 &    $10.937\pm0.259 $ & $91 \pm 2 $ & $6604\pm80$ & $8627\pm298$ & $8.01\pm0.04$ & $0.568\pm0.031$ \\
J100449.541+121559.65 &  $4.497 \pm0.440 $  & $222 \pm 23 $  &  $8974\pm145$ & $9630\pm995$ & $7.85\pm0.11$ & $0.492\pm 0.105$ \\
J111720.801+405954.67 &   $7.448\pm0.215$ & $134 \pm 5 $ &  $10145\pm150$ & $7906\pm330$ & $8.15\pm 0.04$ & $0.660\pm 0.035$ \\
J114722.608+171325.21 &   $6.372\pm0.379 $  & $157 \pm 10 $ &  $9168\pm100$ & $8342\pm531$ & $8.06\pm0.06$ & $0.606\pm0.056$ \\
J124535.626+423824.58 &  $14.116  \pm 0.095 $ & $71 \pm 1 $  & $9295\pm50$ & $7475\pm97$ & $8.23\pm0.01$ & $0.714\pm0.010$ \\
J125507.082+192459.00 &   $8.462 \pm 0.221 $  & $118 \pm 3 $  & $8046\pm50 $& $8526\pm256$ & $8.02\pm0.03$ & $0.580\pm0.027$ \\
J134305.302+270623.98 &$ 5.681\pm 0.336$  & $176 \pm 11 $  &    $9885\pm90$ &  $7387\pm460$ & $8.25\pm0.06$ & $0.727\pm0.049$ \\
J141724.329+494127.85 &   $11.316  \pm 0.075 $ & $88 \pm 1 $ &  $9713\pm70$  & $8295\pm138$ & $8.07\pm0.02$ & $0.614\pm0.015$ \\ 
J151859.717+002839.58 &  $ 5.888 \pm 0.460 $  & $170 \pm 14 $ &   $8244\pm85$ & $7662\pm622$ & $8.19\pm0.07$ & $0.686\pm0.066$ \\
J161704.078+181311.96 &    $ 8.947 \pm 0.193 $  & $112 \pm 2 $ & $7720\pm115$ & $7561\pm283$ & $8.21\pm0.04$ & $0.698\pm0.030$ \\
J230240.032-003021.60 &$ 8.234\pm 0.152$  & $122 \pm 2$  &   $9689\pm60$ &  $8710\pm196$ & $8.00\pm0.02$ & $0.570\pm0.021$ \\
\hline
\end{tabular}
%\end{small}
\tablefoot{ $^\ddagger$ Median distance and symmetric uncertainties reported. No biases, e.g. Lutz-Kelker have been considered.}
\end{table*}

\section{Discussion} \label{s:discussion}

\begin{figure}
\includegraphics[scale=.45]{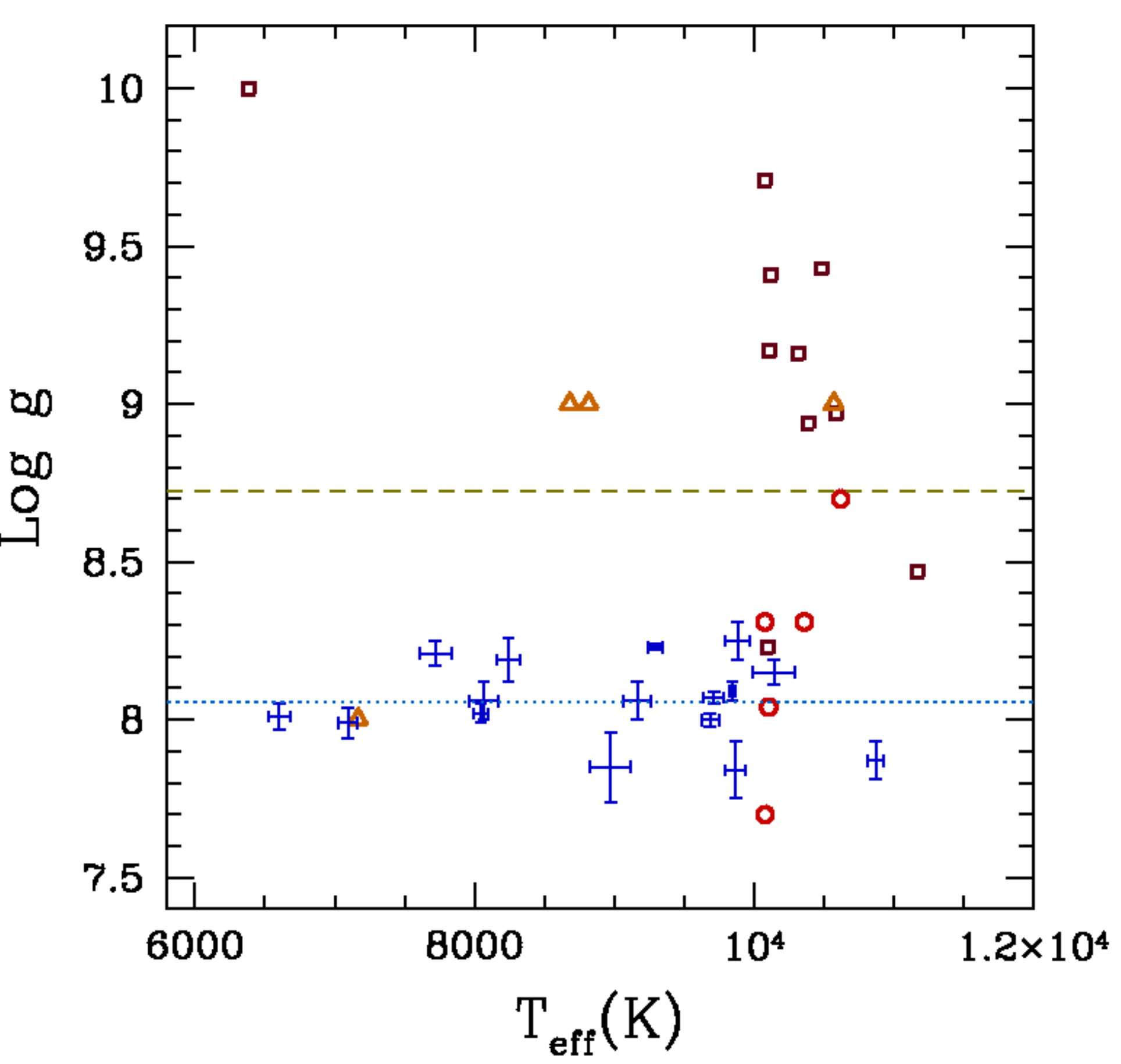}
\caption{Comparsion of $\teff$ and $\logg$ results for the SF18 sample of stars. Blue dots with error bars (this work), open squares \citep{Kleinman:2013}, open triangles \citep{Kepler:2015}, and open circles \citep{Eisenstein:2006}.  \label{fig:tefflogg}}
\end{figure}

Figure~\ref{fig:tefflogg} compares our $\teff$ and $\logg$ determinations, in blue with error bars, with previous results from the literature 
%\citep{mccook:1999, kleinman:2004, eisenstein:2006, kleinman:2013, kepler:2015}. 
\citep{Eisenstein:2006, Kleinman:2013, Kepler:2015}, based on photometry adopting He-pure model atmospheres  without astrometric data. The large scatter in $\logg$ found in previous results stems from the difficulty in the determination for DC stars which lack spectral lines. It is only with the inclusion of Gaia parallaxes that this can be improved substantially. It should also be noted that using He-pure model atmospheres affects the determination of $\teff$, as shown in previous sections.

\begin{figure}
\includegraphics[scale=.19]{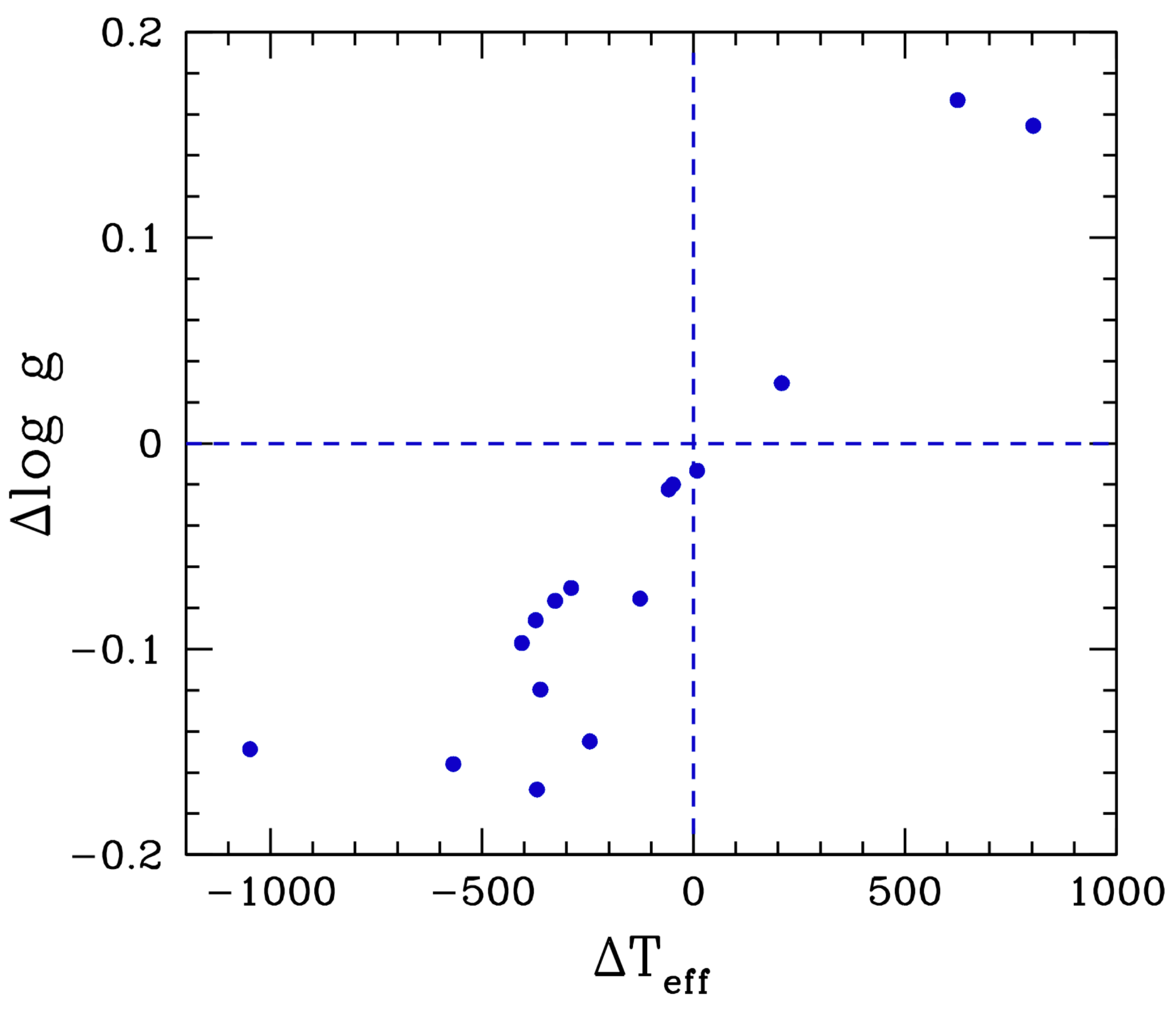}
\caption{$\teff$ and $\logg$  differences in the sense (this work - G19).  \label{fig:gentile}}
\end{figure}

Very recently, \citet[hereafter G19]{gentile:2018} presented a catalogue of white dwarfs with SDSS photometry and Gaia DR2 astrometry. They have determined stellar parameters using either H-pure or He-pure model atmospheres for all stars in the catalogue. Differences between our $\teff$ and $\logg$ determinations with those obtained  with He-pure models by G19 are shown in Figure~\ref{fig:gentile}. The difference in $\teff$ determinations propagates to $\logg$ because a known distance fixes the stellar flux $F_\nu$, so that changes in $\teff$ need to be compensated by changes in the inferred stellar radius. There are two likely  reasons for the difference in the $\teff$ determinations. We have complemented the SDSS $ugriz$ data with Galex UV photometry and used H/He model atmospheres, required to reproduce UV colors.

The mass distribution of our sample has a mean value of
$\langle M_{\rm SF18}\rangle=0.606\,\msun$ and a dispersion $\sigma_{\rm M_{SF18}}=0.075\,\msun$. Results from G19, on the other hand, give $\langle M_{\rm SF18}\rangle=0.651~\msun$ with $\sigma=0.059~\msun$. The difference can be explained because G19 used He-pure models.
When we redo our analysis using He-pure model atmospheres we obtain $\langle M_{\rm SF18}\rangle=0.649~\msun$. Neglecting the presence of small traces of H in the atmospheres can lead to overestimate of mass by about 10\%,  a systematic difference larger than the precision with which the stellar mass is determined. 

We also compare the  mass distribution of our sample with those of stars available in the literature. \citet{koester:2015} has obtained $\langle M_{\rm  DB}\rangle=0.606\,\msun$ with 1-$\sigma$ dispersion between 0.04 and
0.1\,$\msun$ depending on the $\teff$ range analyzed. More recently, using Gaia DR2 parallaxes, \citet{tremblay:2018} has reanalyzed the DB sample from \citet{koester:2015} and found $\langle M_{\rm DB}\rangle=0.580\,\msun$ with a dispersion of 0.087\,$\msun$, and the DB sample from \citet{rolland:2018}, for which they found
$\langle M_{\rm DB}\rangle=0.588\,\msun$ and $\sigma= 0.056\,\msun$.

Finally, the high proximity of these stars to the blackbody spectrum
motivates us to use them as photometric and spectrophotometric
calibrators. Unlike
the use of DA WDs \citep{bohlin:2011, rauch:2016, narayan:2018}, it is unnecessary 
to rely on model atmospheres.
In fact, SF18 used them to verify the zero points of the
photometric system of SDSS across the five-colour bands, showing that
the deviation of the photometric zero points is less than 1\% across the $u$ to $z$ bands. This is
gratifying since the SDSS photometric system is constructed by
concocting several elements and hence it is highly desired to verify
the system, while it is often taken as the standard when one adopts the
AB magnitude system. The SF18 stars also serve as the standard in much wider
range of the spectrum, relative to the optical, from, say, FUV to
several microns in NIR, where the absolute standard is not easily
available.

\section{Conclusions} \label{s:conclusions}

SF18 have identified a group of stars in SDSS with spectral energy
distributions that match almost exactly those of a blackbody. Moreover, they have shown
that these stars can be used as excellent
calibrators for photometric surveys. We confirm the nature of these stars as DCs, and find strong support 
that they are the cool descendants of DB WDs. For this, we use optical and UV photometry and Gaia 
parallaxes to determine the physical parameters 16 of the 17 stars that form the SF18 sample.

We find that these stars are cool He-rich WDs polluted by hydrogen, in the range $-6 < \loghhe < -5.4$, 
having a typical WD mass. The hydrogen abundance has a lower limit imposed by the necessary contribution of 
free electrons so that blackbody colors are reproduced even for the coolest stars in the sample, an effect particularly important when considering UV colors. The 
upper limit is given by the absence of apparent spectral features. The narrow range allowed by these two 
conditions is a possible reason for the paucity of stars with nearly perfect blackbody colors, such 
as those in the SF18 sample, among the much larger set of DC WDs.

The inclusion of Gaia parallaxes allows a refined determination of the surface gravity of the 
stars, otherwise badly constrained due to the lack of spectral lines. We have also shown 
that using He-pure models, as done recently in the literature, leads to systematic offsets in $\logg$ 
determinations that can be traced to systematic differences in $\teff$ determinations. Moreover, using 
He-pure models tends to overestimate the masses of these stars by about 10\%. It is important to note 
that these systematic errors might be a common feature affecting the determination of stellar parameters 
of many, if not all, DC stars for which trace elements are not detected in their spectra but have an 
effect on their photometric properties. The mass distribution for the stars in the SF18 sample agrees 
very well with that of (hotter) DB stars, giving a strong support to the idea that DC stars are their 
cool descendants. 

Finally, we reinforce the possibility, already realized and tested in SF18, that the simplicity of the spectral energy distributions of these stars offers very good potential for calibration of photometric and spectroscopic surveys at the sub-percent level, without the need to rely on stellar model atmospheres.

\begin{acknowledgements}
We thank the anonymous referee for the comments that have helped improving the article.
We would like to thank Max-Planck-Institut f\"ur Astrophysik (MPA)
where this work was initiated. AS is partially supported by the Spanish Government 
(ESP2017-82674-R) and Generalitat de Catalunya (2017-SGR-1131). RDR thanks the support
of the MINCYT (Argentina) through Grant No. PICT 2016-1128. MF thanks
Hans B\"ohringer and late Yasuo Tanaka for the hospitality at the
Max-Planck-Institut f\"ur Extraterrestrische Physik and Eiichiro
Komatsu at MPA, in Garching. He also thanks Alexander von Humboldt
Stiftung for support during his stay in Garching, and Monell
Foundation in Princeton. He received in Tokyo a Grant-in-Aid
(No. 154300000110) from the Ministry of Education.  Kavli IPMU is
supported by World Premier International Research Center Initiative of
the Ministry of Education, Japan. This work has made use of
CosmoHub developed by the Port d'informaco\'o
Cienti\'ifica (PIC), maintained through a collaboration of the
Institut de F\'isica d'Altas Energies (IFAE) and the Centro de
Investigaciones Energ\'eticas, Medioambientales y Tecnol\'ogicas
(CIEMAT), and was partially funded by the "Plan Estatal de
Investigaci\'on Cient\'ifica y T\'ecnica y de Innovaci\'on" program of
the Spanish government.
\end{acknowledgements}

\bibliographystyle{aa}
\bibliography{dbsbbs}
\end{document}